\documentclass{iau} 
\usepackage{graphicx}

\title[JD 11.~~Variation of fundamental constants and white dwarfs] 
{Variation of fundamental constants \\ and white dwarfs }

\author[Susana J. Landau ]   
{Susana J. Landau$^{1,2}$
 }

\affiliation{$^1$Departamento de F\'{\i}sica, Facultad de Ciencias Exactas y Naturales, Universidad de Buenos Aires and IFIBA, 
Ciudad Universitaria - Pab. I, Buenos Aires 1428, Argentina \\ email: {\tt slandau@df.uba.ar} \\[\affilskip]
$^2$CONICET, Godoy Cruz 2290, 1425 Ciudad Aut\'onoma de Buenos Aires, Argentina}

\pubyear{2020}
\volume{357}  
\jname{White Dwarfs as Probes of Fundamental Physics and Tracers of Planetary, Stellar, and Galactic Evolution}
\editors{A.C. Editor, B.D. Editor \& C.E. Editor, eds.}
\begin{document}

\maketitle

\begin{abstract}
Theories that attempt to unify the four fundamental interactions and   alternative theories of gravity  predict  time and/or spatial variation of the fundamental constants of nature. Different versions of these theories  predict different
behaviours for these variations. In consequence, experimental
and observational bounds are an important
tool to check the validity of such proposals.    In this paper, we review  constraints on the possible variation of the fundamental constants from astronomical observations and geophysical experiments designed to  test the constancy of the fundamental constants of nature over different timescales. We also focus on the limits that can be obtained from white dwarfs, which can constrain the variation of the constants with the gravitational potential.
\keywords{ quasars: absorption lines, cosmology: theory,  cosmic microwave background,  supernovae: general , stars: white dwarfs}
\end{abstract}

\firstsection 
\section{Introduction}
\label{intro}
Our present knowledge of the properties of nature is based in two fundamental theories: General Relativity (GR) which describes the gravitational interaction and the Standard Model of Elementary Particles which depicts the electromagnetic, strong and weak interactions. In this way, all physical phenomena at low energies can be described by the solutions of the equations of both theories. However, there is something missing in this picture, namely, there are 20 parameters in these equations that are not given by the theories and must be determined by experiments. These parameters are the ones that  we call the fundamental constants and examples of them are the masses of elementary particles like quarks and leptons;  the gauge coupling constants of the electromagnetic, weak and strong interactions;  the gravitational constant $G_N$, the velocity of light $c$ and  the Higgs vacuum expectation value $<v>$. In turn, the principle of equivalence on which GR is based, requires the invariance of these quantities against changes in position, time and reference system. The gauge coupling constants  of $U(1)$, $SU(2)$ and $SU(3)$ are related to the fine structure
constant $ \alpha $, the QCD energy scale $\Lambda _{QCD}$ and the
Fermi coupling constant $G_F$ through the following equations:
\begin{equation}
\alpha ^{-1}\left( E\right) =\frac 52\alpha _1^{-1}\left( E\right) +\alpha
_2^{-1}\left( E\right)
\end{equation}

\begin{equation}
\Lambda _{QCD} =E\exp \left[ -\frac{2\pi }7\alpha
_3^{-1}\left( E\right) \right]
\end{equation}

\begin{equation}
G_F=\frac{\pi \ \alpha _2\left( M_W\right) }{\sqrt{2}M_W^2}
\end{equation}
where $M_W$ refers to the mass of the W boson. Furthermore, gyromagnetic factors of atoms $g_i$ are also considered fundamental constants and for the purpose of this paper we will consider the proton's mass $m_p$ as a fundamental constant.

Since the Large Number hypothesis was formulated by Dirac in 1937, the variation of the fundamental constants has been the subject of numerous research papers. \cite{Dirac} observed a remarkable coincidence: the dimensionless ratio between the gravitational and the electromagnetic force between a proton and an electron is of order $10^{-39}$ while the time it takes for the light to pass through a hydrogen atom  is just 3 orders of magnitude smaller. This remarkable coincidence led him to formulate the Large Number Hypothesis according to which one or more constants are simple functions of the age of the universe. Furthermore, Dirac proposed that the variation of $G_N$ is of the form: $G_N \sim t^{-1}$. Simultaneously,  \cite{Milne} also considered a possible variation in the $G_N$ but with a different dependence with time : $G_N \sim t$. Following Dirac's proposal, \cite{Teller} suggested that the fine structure constant is also a function of cosmological time and in particular he proposed the following dependence: $\alpha \sim (\log t)^{-1}$. Later, \cite{PS64}, showed that Dirac's proposal cannot explain the  Sun's observed age and luminosity. In an attempt to rescue Dirac's idea, \cite{Gamow} proposed that while $G_N$ remains constant, $\alpha$ varies as a linear function of time. However,  \cite{Dyson} ruled out Gamow's proposal using the abundance of long lived $\beta$ decayers in meteorites. At the same time, \cite{BS67} reached the same conclusion  using  quasar absorption system spectra. The interest in time and spatial variation of fundamental constants got renewed when the attempt to unify the four interactions of nature resulted in the development of multidimensional theories
such as Kaluza-Klein theories (Kaluza 1921, Klein 1926, Marciano 1984), string theories (Maeda 1988, Barr \& Mohapatra 1988, Damour \& Polyakov 1994, Damour et al. 2002) and related brane theories (Youm 2001, Palma et al. 2003). In Kaluza-Klein theories, the variation of the fundamental constants is related to the cosmological evolution of the radius of the extra dimensions, while in the case of superstrings and branes, the variation of the vacuum expectation value of a non-massive scalar field (for example the dilaton in string theories)  is responsible for such variation. 
On the other hand, scalar-tensor theories of gravity are natural frameworks to study the variation of $G_N$. In these theories, a scalar field  couples  to the Ricci scalar and its dynamics provides the physical mechanism for the variation in $G_N$  (Jordan 1959, Brans \& Dicke 1961). More recently, \cite{Moffat2006} proposed an alternative theory of gravity, which attempts to explain  observational data from a large number of astrophysical scenarios without the need to include  dark matter. In the latter, $G_N$ is also allowed to vary with space and/or time.   
Following a different path of research, \cite{Beckenstein} proposed a theoretical framework to study the fine structure constant variability based on general assumptions: covariance, gauge invariance, causality and time-reversal invariance of electromagnetism, as well as the idea that the Planck-Wheeler length $\left(10^{-33}{\rm cm}\right)$ is the shortest scale allowable in any theory. Similar phenomenological frameworks based on first principles were also proposed by other authors (Barrow et al. 2002, Olive \& Pospelov 2002, Chamoun et al. 2001, Barrow \& Magueijo 2005). In these phenomenological frameworks,  the physical mechanism responsible for such   variation  is a scalar field which is added to the theory. 

The experimental research can be grouped into astronomical and local methods. The latter ones include geophysical methods such as the natural nuclear reactor that operated about $1.8\ 10^9$ years ago in Oklo, Gabon, and laboratory measurements such as   detailed comparisons of several atomic clock frequencies with different atomic number. The astronomical methods are
based mainly in the analysis of spectra from high-redshift quasar absorption systems.  Moreover, data from type Ia supernovae also provide limits on the possible variation in $\alpha$. Besides, primoridal nucleosyntesis and the Cosmic Microwave Backdground (CMB) provide constraints on the  the variation of the fundamental constants in the early universe. Moreover, white dwarf spectra and the mass-radius relation in white dwarfs also provide stringent constraints on the variation in $\alpha$ and the proton to electron mass $\mu = \frac{m_p}{m_e} $. 
It was mentioned above that in most of the theories that predict variation of the fundamental constants,   the physical mechanism for such variation is the dynamics of a scalar field.  Near massive objects, like a white dwarf,  the effect of the scalar field can change.  Therefore, constraints on the fundamental constants from white dwarfs, can also be regarded as limits on those variations in terms of a variation in  the gravitational potential.
In such a way, the constraints on $\frac{\Delta \alpha}{\alpha}$ can be translated in terms of a dependence on a dimensionless gravitational potential:

\begin{equation}
 \frac {\Delta\alpha}{\alpha} \simeq  k^{(1)}_{\alpha} \Delta\Phi +  k^{(2)}_{\alpha} (\Delta\Phi)^2 
\label{alfapotential}
\end{equation} 

Similar,  constraints on the possible variation of the  proton to electron mass $\mu$ can be expressed as follows:

\begin{equation}
 \frac {\Delta\mu}{\mu} \simeq  k^{(1)}_{\mu} \Delta\Phi +  k^{(2)}_{\mu} (\Delta\Phi)^2 
\label{mupotential}
\end{equation} 
In Equations \ref{alfapotential} and \ref{mupotential} $\Phi$ refers to the gravitational potential and the coefficients $k^{(i)}_\alpha$, $k^{(i)}_\mu$ can be determined by experimental or observational data. In atomic clock experiments, the difference in the gravitational potential is of order $\Delta \Phi \sim 10^{-10}$, while the difference between the gravitational potential in a white dwarf and the respective one on  Earth is    $\Delta \Phi \sim 10^{-4} -  10^{-5}$. This is the main reason to affirm that white dwarf spectra  are the ideal probe for a relationship between the fundamental constants like $\alpha$ and $\mu$  and strong gravitational fields. In this article, we will describe bounds on the fundamental constants from geophysical experiments and astronomical observations with a focus on those that can be obtained from white dwarfs. In section \ref{geo} we  describe the bounds from geophysical data like the Oklo nuclear reactor and atomic clocks. We also discuss the bounds on the relation between the fundamental constants and the gravitational potential that can be obtained with this method.
In section \ref{quasar} we review the bounds from quasar absorption systems and discuss the status of the  claimed variation in $\alpha$ from the Many Multiplet Method. Besides, bounds from primordial nucleosynthesis and the Cosmic Microwave Background are discussed in Section \ref{earlyuniverse}. Furthermore, bounds from type Ia supernovae data are described in Section \ref{supernovae}. A detailed discussion on the bounds that can be obtained from white dwarf data on the possible variation of $\alpha$, $\mu$ and $G_N$ is described in Section \ref{wd}.  We also provide a description on the bounds on $G_N$ that can be obtained from the Lunar Laser Ranging experiment (see Section \ref{llr}) and helioseismology (see Section \ref{helio}). Finally, in Section \ref{conclusions} we discuss our conclusions.
  
\section{Bounds from geophysical data}
\label{geo}
\subsection{The Oklo Phenomenon}
One of the most stringent limits on time variation of the fine
structure constant $\alpha$ follows from an analysis of isotope ratios
in the natural uranium fission reactor that operated $1.8\times 10^9$
years ago at the present day site of the Oklo mine in Gabon,
Africa. From an analysis of nuclear and geochemical data, the
operating conditions of the reactor could be reconstructed and the
thermal neutron capture cross sections of several nuclear species
measured. In particular, a shift in the lowest lying resonance energy level
in $^{149}{\rm Sm}: \Delta = E_r^{149{\rm(Oklo)}} -
E_r^{149{\rm(now)}}$ can be derived from a shift in the neutron capture cross section of the same nucleus (Damour \& Dyson 1996).
The shift in $\Delta $ can be translated  into a bound on a
possible difference between between the values of  $\alpha $ and $G_F$ during the Oklo phenomenon
and their value now:
\begin{equation}
\Delta =10^6 {\rm eV} \frac{\Delta \alpha }\alpha + 5.6 {\rm eV} \frac{\Delta G_F }{G_F}
\end{equation}
Several authors have analyzed the Oklo data to put constraints on the possible variation of the fundamental constants (Petrov et al. 2006, Davis \& Hamdan 2015). The most stringent limit was established by \cite{DH15}:

\begin{equation}
\frac{\Delta \alpha }\alpha < 1.1 \times 10^{-8}
\end{equation}

\subsection{Atomic Clocks}
\label{atomicclocks}
Atomic clocks provide the most stringent bounds on the variation of fundamental constants. The method consists of comparing atomic clock frequencies of  different transitions, which in turn have different dependence with the fundamental constants. For example, the dependence of a hyperfine transition frequency with the fundamental constants can be expressed as follows:
\begin{equation}
\nu_{\rm Hyp} \sim  \alpha^2 g_i \frac{m_e}{m_p} R_{\infty} c  F_{REL}(\alpha Z)
\nonumber
\end{equation}
where $g_i$ is gyromagnetic factor, $R_\infty$ is Rydberg's constant, $m_e$ and $m_p$ are the electron's and proton's mass respectively and  $F_{REL}$ is the  relativistic contribution to the energy. In such way, clocks based on
hyperfine transitions in alkali atoms with different atomic number $Z$
can be used to set bounds on $\alpha^k \frac{\mu_{A_1}}{\mu_{A_2}}$
where $k$ depends on the frequencies measured and $\mu_{A_i}$ refers
to the nuclear magnetic moment of each atom (Prestage et al. 1995, Marion et al. 2003, Gu{\'e}na et al. 2012). 

On the other hand, an optical transition frequency
has the following dependence on $\alpha$:
\begin{equation}   
\nu_{opt} \sim R_{\infty} B F_i(\alpha)
\end{equation}
where $B$ is a numerical constant assumed not to vary in time and
$F_i(\alpha)$ is a dimensionless function of $\alpha$ that takes into
account level shifts due to relativistic effects. Thus, comparing an
optical transition frequency with a hyperfine transition frequency
can be used to set bound on $\alpha^k \frac{m_e}{m_p}
\frac{\mu_A}{\mu_B}$ (Le Targat et al. 2013, Huntemann et al. 2014, Falke
et al. 2014). Furthermore, the most stringent bound on $\alpha$ variation using this method was obtained by \cite{Rosenband08} and it follows from a comparison between  two optical transition frequencies.  On the other hand, the energy difference between two adjacent rotational levels in a diatomic molecule is proportional to $\frac{1}{M r^2}$, $r$ being the bond length and $M$ the reduced mass. Morevover, the vibrational transition of the same molecule has, in
first approximation, a $\sqrt{M}$ dependence. In this way, it follows that the vibro-rotational molecular transitions are proportional to  $\frac{1}{\sqrt{M}}$. Therefore, comparing  vibro-rotational transitions with an hyperfine transition gives information about the variaton of  $g_{\rm hyp} \sqrt{\frac{m_e}{m_p}} \alpha^k$.  \cite{Shel08} used this latter method to  compare  the vibro-rotational transition of the $SF_6$ molecule with the $^{133}$Cs hyperfine transition. Their results   can also be used to constrain $k_{\mu}^{(1)}$ from Eq.\ref{mupotential}. In this experiment,  the difference in the gravitational potential in the Earth due to the Sun between aphelion and perihelion was estimated to be  $\Delta \Phi = 10^{-10}$,  and therefore the  bound on $\frac{\Delta \mu}{\mu}$ yieds:

\begin{equation}
k^{(1)}_{\mu} < 4 \times 10^{-4}
\nonumber
\end{equation}
On the other hand, comparison of radio-frequency transitions between nearly degenerate, opposite parity excited states in atoms allows us to establish bounds on the possible variation in $\alpha$ due  to the large relativistic corrections of opposite sign for the opposite-parity levels. This method was used by \cite{Leefer13} to establish a bound in $\alpha$ from the comparison of  frequency transitions in two isotopes of
atomic dysprosium (Dy). This bounds can also be used to set constraints on $k_{\alpha}^{(1)}$ in Equation \ref{alfapotential}. In fact, since the estimated  difference in the gravitational potential  between aphelion and perihelion for this experiment is  $\Delta \Phi = 3.3 \times 10^{-10}$,  the bound on $\frac{\Delta \alpha}{\alpha}$ yieds:
\begin{equation}
k^{(1)}_{\alpha} = (-5.5 \pm 5.2) \times 10^{-7}
\nonumber
\end{equation}
Finally, combining all available laboratory data   \cite{Godun14} obtain:
\begin{eqnarray}
\frac{\dot\alpha}{\alpha}=(0.7 \pm 2.1) \times 10^{-17} {\rm yr}^{-1}  \\ 
\frac{\dot\mu}{\mu}=(0.2 \pm 1.1) \times 10^{-16} {\rm yr}^{-1}   
\end{eqnarray}
\section{Bounds from Quasar Absorption Systems}
\label{quasar}
Quasar absorption systems present ideal laboratories in which to test for possible variations of the fundamental constants.  
The relationship between the wavelength observed in quasar spectra ($\lambda _{\rm obs}$) and the ones measured in the laboratory ($\lambda _{\rm lab}$)  can be expressed:
\begin{equation}
\lambda _{\rm obs}=\lambda _{\rm lab}\left( 1+z\right) .
\end{equation}
Several methods  have been developed to test the possible variation of the fundamental constants, among them, it is important to mention the following: i) Alkali Doublet Method, ii) Many Multiplet Method, iii) Comparison of molecular spectra with  laboratory spectra, iv) Comparison of radio spectra with optical spectra, v) Comparison of radio spectra with molecular spectra and vi) Conjugate Lines Method. Some of them such as the Akali Doublet Method rely on the comparison between the observed  wavelengths in the quasar  and the one measured in the laboratory. Others, are based on the comparison of absorption redshifts due to different transitions that happen in the same absorption cloud. We will not discuss all  methods in this article. We will focus on i) the Many Multiplet Method which is so far the method that allows us to place the more stringent constraints on $\frac{\Delta \alpha}{\alpha}$from quasar spectra and ii) the comparison between molecular and optical spectra which provides stringent constraints on the variation of $\mu$.

The Many Multiplet Method  was proposed by \cite{Webb99} and relies on the comparison of  transitions with different atomic masses in the
same absorption cloud. In short, the Many Multiplet Method consist of adjusting the Voigt profiles of 
several absorption lines, 
including besides the usual fit parameters: column density, Doppler
width, redshift, a possible variation of the fine structure constant. In this way, the method allows us to gain an order of magnitude in
sensibility with respect to previously reported data. Using this method and data provided by the Keck telescope, \cite{Webb99} claimed a detection of a variation in $\alpha$. However, an independent analysis performed with observations made with UVES at the Very Large Telescope (VLT) provided null results (Srianand et al. 2004). Contrary to previous results,  a third analysis, this time with  VLT/UVES archival data also indicated a variation in $\alpha$,
but now with $\alpha$ increasing with redshift (Webb et al. 2011, King et al. 2012). These results, led the authors to suggest a spatial dipole-type variation in $\alpha$.
The weighted mean of the 293 measurements obtained with data from both Keck and VLT telescopes by the group of Webb et al  at $0.3 < z < 3.1$ is:
\begin{equation}
\left(\frac{\Delta\alpha}{\alpha}\right)=-2.16\pm0.86\, \times 10^{-6},.
\end{equation}
On the other hand, more recently,  a reanalysis of systematic errors with new techniques showed that there is no compelling evidence for any variation in $\alpha$ from quasar data (Whitmore \& Murphy 2015).
However, it should be noted that in all those mentioned analyses the data acquisition procedures were far from ideal, in particular, regarding the key issue of wavelength calibration.
Trying to confirm the above mentioned results was the main motivation for the ESO UVES Large Program, for which key improvements in the data acquisition procedures were implemented (Bonifacio et al. 2014).  Furtheremore, other dedicated measurements of the variation in $\alpha$ with the quasar method were performed recently, by several authors (Agafonova et al. 2011, Bainbridge \&
Webb 2017, Evans et al. 2014, Kotus et al. 2017, Molaro et al. 2013, Songaila \&
Cowie 2014). In particular  \cite{Murphy16} developed a  method to test a variation in $\alpha$ which is not influenced by long-range distortions. The weighted mean of all recent dedicated  measurements which span the redshift range $1 < z < 2.4$  is (Martins 2017): 
\begin{equation}
\left(\frac{\Delta\alpha}{\alpha}\right)=-0.64\pm0.65\, \times 10^{-6},
\end{equation}
showing no variation in $\alpha$. Nevertheless, \cite{Murphy16} reached the conclusion  that their quasar sample is too small to rule out the dipole model. On the other hand, it should be noted that all above mentioned measurements
were performed with spectrographs such as UVES, HARPS or Keck-HIRES, which are far from optimal for this type of measurement. Therefore,  more precise measurements using the new generation of high-resolution spectrograph, like ESPRESSO for the VLT and E-ELT-HIRES for the E-ELT, are expected to  improve significantly the precision of the data.

\cite{VL93} developed a method for
constraining the variation in $\mu = \frac{m_p}{m_e} $ which is  based on the fact that wavelengths of electron-vibro-rotational lines depend on the reduced
mass of the molecules, with different dependence for different transitions. In such way, it is possible to distinguish the cosmological redshift of a line from the shift caused by a variation
in $\mu$. The wavelength of a molecular line at redshift $z_{\rm abs}$ can be expressed as:
\begin{equation}
\lambda_i = \lambda_i^{\rm lab} \left(1 + z_{\rm abs} \right)\left(1 + K_i  \frac{\Delta \mu}{\mu}\right)
\nonumber
\end{equation}
where $\lambda_i^{\rm lab}$ is the wavelength measured in the laboratory and $K_i$ is a coefficient for the molecular band. Several authors have used this method to place stringent constraints on the variation in $\mu$ at $0.6< z < 4.2$ (Kanekar 2011, Bagdonaite et al. 2013, van Weerdenburg et al. 2011, Rahmani et al. 2013, Dapr{\`a} et al. 2015, Albornoz V{\'a}squez
et al. 2014, Bagdonaite et al. 2015). Measurements at low redshift were obtained with radio/mm observations and its weighed mean yields $\frac{\Delta \mu}{\mu} = (-0.24 \pm 0.09) \times 10^{-6}$ (Martins 2017) while the high redshift sample proceeds from UV/optical observations with weighed mean  $\frac{\Delta \mu}{\mu} = (2.9 \pm 1.9) \times 10^{-6}$  (In table \ref{tablemu} we show the weighed mean of the complete data set). There is weak evidence for a variation in $\mu$ from both samples, even though it should be stressed that the sign of the variation is different at high and low redshift. Besides, \cite{Bagdonaite13b} obtain  the most stringent limit  from observations of metanol transitions at $z=0.89$ :$\frac{\Delta \mu}{\mu} = (-1 \pm 1.8) \times 10^{-7}$.

\section{Bounds from the early universe}
\label{earlyuniverse}
\subsection{Cosmic Microwave Background}

The Cosmic Microwave Background (CMB) radiation is one of the best tools to study the early universe because it provides   information about the physical conditions in the Universe just before decoupling of matter and radiation. Possible variation in $\alpha$ and the electron mass $m_e$ affect the Thompson scattering cross section $\sigma _T=\frac{8\pi \ \hbar ^2c^2}{3\  m_e^2} \alpha^2$ and the ionization fraction. The main effect in the ionization fraction can be expressed as follows: 
\begin{equation}
x_e\simeq \left( \frac{m_e}{k_BT}\right)^{\frac 32}\exp -\left( \frac{B_1}{k_BT}\right) 
\end{equation}
where $B_1=\frac{\alpha ^2 m_e c^2}{2}$ is the Hydrogen binding energy. The effect of a possible variation in $\alpha$ and/or $m_e$ on the CMB doppler peaks is a shift in the position of the peaks and modification in the height of the peaks. From the comparison of the theoretical prediction with observational data obtained by the \cite{Planck15}, bounds on the possible variation in $\alpha$ and/or $m_e$ can be obtained. Assuming indepedent variations in $\alpha$ and $m_e$, \cite{HC18} obtain:
\begin{equation}
\left(\frac{\Delta\alpha}{\alpha}\right)=(0.7\pm 2.5)\, \times 10^{-3} \,\,\, \,{\rm and }\,\,\, \,\left(\frac{\Delta m_e}{m_{e0}}\right)=(3.9 \pm 7.4)\, \times 10^{-3} 
\end{equation}
where the last bound was obtained considering also Baryon Acoustic Oscilation (BAO) data in the statistical analysis. On the other, hand if  joint variations of $\alpha$ and $m_e$ are assumed and BAO data are considered in the analysis, the  limits obtained are:
\begin{equation}
\left(\frac{\Delta\alpha}{\alpha}\right)=(1.1\pm 2.6)\, \times 10^{-3}  \,\,\, \,{\rm and }\,\,\, \,\left(\frac{\Delta m_e}{m_{e0}}\right)=(5.6 \pm 8)\, \times 10^{-3} 
\end{equation}

\subsection{Big Bang Nucleosynthesis (BBN)}
\label{nucleo}
The study of the light nuclei production during the first three minutes of the universe is another important tool to study the early universe.
In short, the abundances of light nuclei depend on the following physical parameters: freeze-out time of the weak interactions, neutron-proton mass diference, binding energies of the light nuclei and cross sections of the reactions producing $D$, $^4${\rm He}, $^7${\rm Li}, $^6${\rm Li}, Tritium, $^3${\rm He}, and $^7${\rm Be}, which, in turn are  functions of the fundamental constants $\alpha$, the Higgs vacuum expectation value $<v>$ and $G_N$. It should be stressed that unlike other methods described in this paper, it is necessary to asume a theoretical model for the strong interaction within this analysis.  \cite{MC17} considered the   Argonne potential for the nucleon-nucleon interaction. From the comparison with the observed abundances of $D$, $^4${\rm He} and  $^7${\rm Li} they obtained :
\begin{equation}
\frac{\Delta \alpha}{\alpha}= -0.022 \pm 0.006   \,\,\,\, \,\,{\rm and} \,\,\,\, \,\, \frac{\Delta <v> }{<v>}=0.042\pm 0.01
\end{equation}
whereas if the Bonn potential is assumed, the analysis yielded:
\begin{equation}
\frac{\Delta \alpha}{\alpha}= -0.022 ^{+0.003}_{+0.004}   \,\,\,\, \,\,{\rm and} \,\,\,\, \,\, \frac{\Delta <v> }{<v>}=0.036\pm 0.007
\end{equation}
On the other hand, the value of $G_N$  determines the expansion rate of the universe and thus, the relevant time scales for the weak and nuclear reactions. As a consequence, assuming a variation in $G_N$ at the time of the BBN, translates in turn  into a variation of the light element abundances with respect to the ones predicted by the standard cosmological model. \cite{copy04} and \cite{Bambi05} compared the theoretical predictions for the light nuclei abundances with the present abundances of  {\rm D}, $^4${\rm He} and $^7${\rm Li} and established the following bound:
\begin{equation}
\frac{\dot G}{G}= (0.05 \pm 0.35) \times 10^{-12}  {\rm yr}^{-1}
\end{equation}

\section{Bounds from supernovae type Ia}
\label{supernovae}
Type Ia supernovae (SNe Ia)  are among the most energetic and interesting phenomena in our universe. Furthermore,  the spectra and light
curves of normal SNe Ia are very homogeneous in such a way that they are considered one of the best standard candles known today. All this makes them suitable astronomical objects to test the possible variation of fundamental constants. The homogeneity of the light curve is basically caused by  the homogeneity of the nickel mass produced during the supernova outburst. This effect is primary determined by the value of the Chandrasekhar mass which in turn depends on the value of the fundamental constants $G_N$, $\alpha$ or the velocity of light $c$ as follows:
\begin{equation}
M_{\rm ch} \sim  \left(\frac{\hbar c}{G_N}\right)^{3/2} \sim  \left(\frac{e^2}{\alpha^2 G_N}\right)^{3/2}
\end{equation}
In addition, the possible variation of the fundamental constants with  cosmological time would also change the cosmic evolution and therefore affect the luminosity distance relation. \cite{Gazta02}  performed an analysis allowing only for the variation in $G_N$ and showed that the latter is typically several times smaller than the
change produced by the corresponding variation of the Chandrasekhar mass. Furthermore,  these authors used data from the Supernova Cosmology Project to put the following bound:
\begin{equation}
\frac{\dot G_N}{G_N} < 10^{-11} {\rm yr}^{-1}
\end{equation}
\cite{Garciaberro06} considered that the variation in   $G_N$ follows the predictions of scalar-tensor theories and  reached similar conclusions.
On the other hand, \cite{Kraiselburd2015} analised the dependence of type Ia supernovae  explosions on $\alpha$, including both the change in the Chandrasekhar mass and the the dependence of the mean opacity of the expanding supernovae photosphere. Furthermore, motivated by the results of the quasar data, they considered a spatial dipole-type variation for $\alpha$. From the comparison of supernovae data from the Union 2.1 compilation, they obtained:
\begin{equation}
\frac{\Delta \alpha}{\alpha} = (2.1 \pm 0.8) \times 10^{-2}
\end{equation}
In a posterior work, \cite{Negrelli2018} considered the variation in $c$ as the source of the variation in the Chandrasekhar mass. In addition, they  also included the change of the energy release during the explosion in their analysis. Comparing the theoretical predictions for a spatial dipole-type variation in $c$, with the Union 2.1 and JLA  supernovae type Ia data, they  established:
\begin{equation}
\frac{\Delta c }{c}= (-3.7 \pm 0.8) \times 10^{-2}
\end{equation}
\section{Bounds from White Dwarfs}
\label{wd}
\subsection{Bounds on $\alpha$ and $\mu$}

\begin{figure}[b]
\begin{center}
 \includegraphics[width=3in]{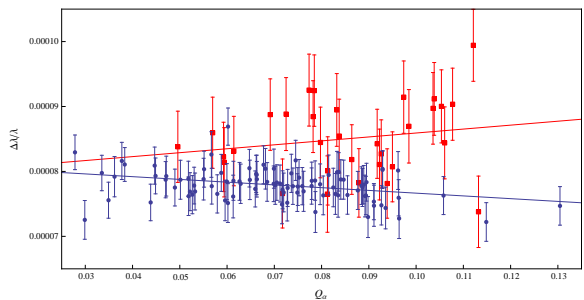} 

 \caption{$\frac{\Delta \lambda}{\lambda}$ vs. $Q_\alpha$ for transitions in FeV (blue circles) and NiV (red squares). The slope of the lines give $\frac{\Delta \alpha}{\alpha} =
(4.2 \pm 1.6) \times 10^{−5}$ for FeV and $ \frac{\Delta \alpha}{\alpha}= (-6.1 \pm 5.8) \times 10^{−5}$ for  NiV, respectively.  Reprinted with permission from Physical Review Letters 111, 010801  (2013). Copyright(2013) by the American Physical Society}
   \label{fig1}
\end{center}
\end{figure}

Hot white dwarf stars are ideal laboratories to probe for a relationship between the fundamental constants  and strong gravitational
fields. Hot white dwarfs with masses comparable to the sun and radii comparable to Earth generate
strong gravitational fields and are typically bright with numerous absorption lines. The relationship between the laboratory wavelengths ($\lambda _{\rm lab}$) and those observed near
a white dwarf ($\lambda _{\rm WD}$) is :
\begin{equation}
\frac{\Delta \lambda }{\lambda}=\frac{\lambda_{\rm WD} -\lambda _{\rm lab}}{\lambda
_{\rm lab}}=z_{\rm WD} -Q_{\alpha }\frac{\Delta \alpha }{\alpha }(1+z_{\rm WD})\,,  \label{lambda}
\end{equation}%
where $Q_{\alpha }=2q/\omega _{0}$ is the relative sensitivity of the
transition frequency to a variation in $\alpha $. \cite{Berengut13} observed the wavelength shift in 96 quadruply ionized iron and 32 quadruply ionized nickel absorption features from the white dwarf star G191-B2B and derived separate limits for each metal. In all measurements line calibration was the expected culprit for the observed signal. Both results (see Figure \ref{fig1}) are inconsistent with each other at the 1.6 sigma level, the likely reason being related to uncertainties in laboratory wavelength measurements and systematics in the wavelength calibration. More recently, \cite{Bainbridge17} presented preliminary results of a similar analysis incorporating improvements such as; i) the use of new laboratory wavelengths; ii) measurements of a sample of objects rather than a single object ; and iii) the use of robust techniques from quasar absorption systems  (the Many Multiplet method) in the data analysis method. Therefore, future measurements with improved data analysis methods could provide more stringent bounds.

On the other hand, a possibe variation in $\mu$ would be manifested as shifts in the observed wavelengths of molecular hydrogen ($H_2$)  when compared to laboratory
wavelengths as follows

\begin{equation}
 \frac{\lambda_i^{WD}}{\lambda_i^{\rm lab}} =  (1+z_{WD})(1+ \frac{\Delta \mu}{ \mu}K_i)
\nonumber
\end{equation}
where $\lambda_i^{WD}$ represents the $H_2$ transition wavelength observed in white dwarf spectra, $\lambda_i^{\rm lab}$ is a corresponding wavelength measured in the laboratory and $K_i$ are the sensitivity coefficients. \cite{Bagdonaite14} applied this method to the spectrum of the white dwarf star  G29$-$38 with a potential of $2 \times 10^4 \Phi_{\rm earth}$ to obtain:  
\begin{equation}
\Delta\mu/\mu=(-5.8\pm3.8_{\rm stat}\pm 0.3_{\rm sys})\times10^{-5}
\end{equation}
while for the white dwarf  GD133 with a potential of $10^4 \Phi_{\rm earth}$ the resulting bound was:
\begin{equation}
\Delta\mu/\mu=(-2.7\pm4.7_{\rm stat}\pm 0.2_{\rm sys})\times10^{-5}
\nonumber
\end{equation}
We have discussed in Section \ref{intro} the theoretical motivation for considering varying fundamental constants and the inclusion of scalar fields as the physical mechanism for that variation in most theoretical models. Therefore,   constraints on $\Delta\mu/\mu$ can be interpreted in terms of a dependence on a dimensionless gravitational potential. Applying Eq.\ref{mupotential} to the above bounds yields the following limit :
\begin{equation}
k^{(2)}_{\mu} < 1 \times 10^{3}
\end{equation}
which is several orders of magnitude more stringent than the ones obtained from Earth-based experiments. And we remember the reader that the constraint on $k^{(1)}_{\mu}$ is obtained from atomic clock experiments (see Section \ref{atomicclocks}).

\begin{figure}[b]
\begin{center}
 \includegraphics[width=3in]{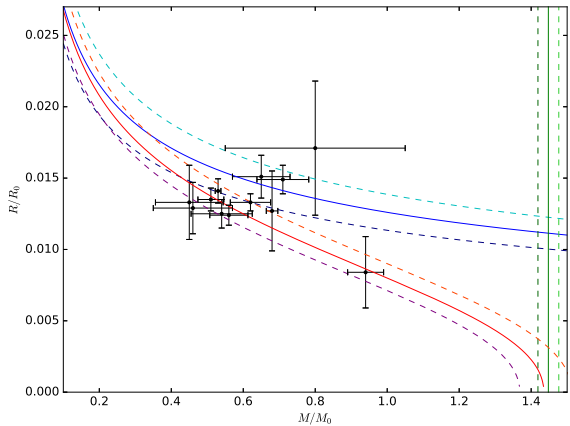} 
 \caption{Solid red line for the standard model, with
the nearby darker and lighter dashed lines corresponding to $\beta \pm = 0.01$. Solid blue line  shows the nonrelativistic
limit of the polytropic model, and darker and lighter dashed lines
for  $x = \pm 0.1$. $x= \left[ \frac{4}{3}R+\frac{5}{6}(1+S) \right]  \frac{\Delta \alpha}{\alpha}$. The black points with error bars
correspond to the data.  $R_0$ and $M_0$ refer to the Sun's radius and mass respectively. Reprinted with permission from Physical Review D96, 083012 (2017). Copyright(2017) by the American Physical Society }
   \label{fig2}
\end{center}
\end{figure}

On the other hand, \cite{Magano17} obtained bounds on the possible variation of the fundamental constants  from the mass-radius relation in white dwarfs. Unlike the analyses performed for other bounds described in this paper\footnote{with the exception of primordial nucleosynthesis (see Section \ref{nucleo} where a model for the strong interactions has to be assumed)}, it is necessary to assume a theoretical model for the relation between the variation of the electron and nucleon masses and $\alpha$. For this, the authors assume the expressions given by the generic class of unification models proposed by \cite{Coc07}:

\begin{equation}
\frac{\Delta m_e}{m_e}=\frac{1}{2}(1+S)\frac{\Delta\alpha}{\alpha}
\end{equation}
\begin{equation}
\frac{\Delta m_N}{m_N}=[0.8R+0.2(1+S)]\frac{\Delta\alpha}{\alpha}\,
\end{equation}
where $m_N$ refers to the nucleon masses, and  $R$ and $S$ depend of the specific unification model considered. Using these relations, the mass continuity and hidrostatic equations can be expressed for the case of varying constants as follows:
\begin{eqnarray}
\frac{dm'}{dr}&=&m_0 r^2 {x_F}^3\\
\frac{dx_F}{dr}&=&-K_1 (1+\beta) \,\,\frac{m'}{r^2} \frac{\sqrt{1+x_F^2}}{x_F}\\
m&=&K_2 (1+\gamma)\,\,m'
\end{eqnarray}
where $K_1= \frac{16}{3 \pi (2q)^2} \frac{R_{\odot}}{m_0} \frac{c^3}{G \hslash^3} \alpha_e \alpha_N $, $K_2= \frac{8}{3 \pi (2q)8} \frac{{R_{\odot}}^3}{M_{\odot}m_0} \frac{c^5}{G^2 \hslash} {\alpha_e}^{\frac{3}{2}} {\alpha_N}^{\frac{1}{2}}$, $\alpha_i=\frac{Gm^2_i}{\hbar c}$, $x_F=\frac{p_F}{m_e c}$, $p_F$ is the Fermi momentum, $m_0$ is a dimensionless constant and $q$ is the number of electrons per nucleon. In the above equations, $\beta$ and $\gamma$ enclose the dependence of the above equations with the fundamental constants and are given by the following expressions:
\begin{eqnarray}
\beta= \left[ \frac{9}{5}R+\frac{8}{5}(1+S) \right] \frac{\Delta \alpha}{\alpha} \:,
\nonumber 
\gamma= \left[ \frac{4}{5}R+\frac{23}{10}(1+S) \right] \frac{\Delta \alpha}{\alpha}\:.
\nonumber
\end{eqnarray}
Results of the numerical integration of the above equations from the original work are shown in Figure \ref{fig2}, for the cases: i)$\beta=0$ and $\gamma=0$ (standard model), ii) $\beta = \pm 0.01$
and iii)a more simple polytropic model; toghether with present data for the mass-radius relation. Next, a statistical analysis was performed to estimate bounds on $\alpha$ from observational data. For each star $i$ in the catalog, \cite{Magano17} chose a value of $M_{\star}$ to minimize the following quantity:
\begin{equation}
{\chi_i}^2 (M_{\star}) = \frac{{(M_{\star}-M_i)}^2}{\sigma_{M,i}^2}+\frac{{(R_{th}(M_{\star})-R_i)}^2}{\sigma_{R,i}^2} \:,
\end{equation}
where $M_i$, $\sigma_{M,i}$, $R_i$, and $\sigma_{R,i}$ are the mass and radius of the $i$th star and their respective uncertainties, and $R_{th}(M)$ is the theoretical prediction. Thus, the total total value of 
$\chi^2$ to be calculated was  $\chi^2 = \sum_{i=1}^{N} {\chi_i}^2({\hat{M_i}})$ with $\hat {M_i}$ as the value of $M$ that minimizes the corresponding $\chi_i^2$. From the statistical analysis, the following constraints were obtained:
\begin{eqnarray}
\beta=0.012 \pm 0.032 \:, \\
\gamma=0.006 \pm 0.060\:
\end{eqnarray}
If the typical values suggested in Coc et al 2007 were assumed: $R\sim30$ and $S\sim160$, allowing a 10$\%$ uncertainty in each of them, the following bound was established:
\begin{equation}
\frac{\Delta \alpha}{\alpha}=(2.7\pm9.1) \times 10^{-5} \, \nonumber
\end{equation}
Finally, it should be noted, that improvements in mass and radius measurements, such as those expected from  the Gaia satellite,  will allow us to obtain more stringent constraints. 

\subsection{Bounds on $G_N$}

White dwarfs can also be used to put independent limits on a possible variation in $G_N$. The most important reason for this is that white dwarfs have very long
evolutionary timescales. A possible variation of $G_N$ over cosmological ages changes the temperature in the central regions of white dwarf progenitors. As a consequence, the thermonuclear rates, luminosities and main sequences lifetimes of these stars are also modified. Furthermore,  white dwarf cooling times are also affected by a change in $G_N$. \cite{Garciaberro11} calculated the effects of a varying $G_N$ on the main sequence ages using a stellar evolutionary code. Furthermore, employing modified white dwarf cooling ages in order to account for a varying $G_N$,  they built the white dwarf luminosity function for the old, metal-rich Galactic open cluster NGC 6791.  Comparing the theoretical predictions for a varying $G_N$ with the observational color-magnitude diagram and the white dwarf luminosity function, and using the distance modulus measured with eclipsing binaries, the following bound was obtained:

\begin{equation}
\frac{\dot G_N}{G_N} < -1.8 \times 10^{-12} {\rm yr}^{-1}
\end{equation}

On the other hand, pulsational properties of white dwarfs can be used to constrain a possible variation in $G_N$. \cite{Corsico13}  compared the theoretical
rates of period change of white dwarfs including the effects of a running $G_N$ with the measured rates of change of the periods of two well studied pulsating white dwarfs, G117−B15A
and R548. For this, they used a stellar evolutionary code and a pulsational code to compute pulsational properties of variable white dwarfs. Results show that the pulsation periods do not  change in a significant way, but the rates of period change are strongly affected when a varying $G_N$ is assumed. Moreover, they obtain the following bound

\begin{equation}
\frac{\dot G_N}{G_N} < - 1.3 \times 10^{-10} {\rm yr}^{-1}
\end{equation}
which is less stringent that the one obtained from the white dwarf luminosity function.
\section{Bounds from the Lunar Laser Ranging experiment}
\label{llr}
The Lunar Laser Ranging (LLR)  experiment, has  provided high-precision values of the Earth-Moon distance,  through measurements of round
trip travel times of laser pulses between stations on the Earth and retroreflectors on the Moon. 
Results of this experiment have been used to test General Relativity, the Strong Equivalence Principle, metric parameters, preferred frame effects and the temporal variation of the gravitational constant $G_N$. \cite{Hofmann10}  updated the Institut fur Erdmessung (IfE) LLR model taking the effect of a fluid lunar core into consideration. In this way, they reduced the uncertainty in the current variation in $G_N$ by a factor of 2, obtaining:
\begin{equation}
\frac{\dot G_N}{G_N}= (0.7 \pm 3.8) \times 10^{-13} {\rm yr}^{-1}
\end{equation}

\section{Bounds from Helioseismology}
\label{helio}
\cite{Guenther98} considered solar models including the possible variation of the gravitational constant $G_N$ during the solar lifetime. Furthermore, they compared the p-mode oscilation spectra of those models with observations  from the Global Oscillation Network Group (GONG) instrument and Birmingham Solar Oscillation Network (BiSON). As a result of their research, they obtain the following bound:

\begin{equation}
\frac{\dot G_N}{G_N} < 1.6 \times 10^{-12} {\rm yr}^{-1}
\end{equation}

\section{Summary and Conclusions}
\label{conclusions}

In this paper we have described the most important present bounds on the possible variaton of the fundamental constants $\alpha$, $\mu$ and $G_N$. We also have mentioned the dependence of some observables on other fundamental constants such as the velocity of light $c$, the Fermi constant $G_F$ and the Higgs vaccum expectation value $<v>$. In order to have a comprehensive picture,  Tables \ref{tablealpha}, \ref{tablemu} and \ref{tableG} show a summary of the bounds that can be obtained with each of the methods described in this review, together with the time interval for which the change in the fundamental constants was measured. It is important to stress, that bounds on the variation in $\alpha$ and $\mu$ from white dwarf spectra, are constraints of changes  with the gravitational potential (which could also be regarded as a spatial variation), but they do not refer under any point of view to a time variation of fundamental constants. Finally, we would like to emphasize, the most important aspects that were discussed in this review: 
\begin{enumerate}
\item There is no evidence for a time or space variation of the fundamental constants. 
\item The most stringent limits on variation of the fundamental constants are provided by atomic clocks experiments:
\item White dwarf stars are  ideal laboratories to test for a relationship between the fine-structure
constant or the proton-to-electron-mass and  strong gravitational fields. Current constraints are derived from molecular hydrogen measurements and the mass-radius relationship  while future measurements of ${\rm Fe}$ V  and ${\rm Ni}$ V spectra in white dwarfs could provide more stringent bounds. 
\end{enumerate}

\begin{table}[!ht]
\begin{center}
\renewcommand{\arraystretch}{1.3}
\begin{tabular}{|l|l|l|l|}
\hline
Source & $\frac{\Delta \alpha}{\alpha} \pm \sigma$ &  $\Delta t$ ({\rm yr}) & Reference  \\
\hline
Atomic Clocks  &$ (2.1 \pm 6.3) \times 10^{-17}$ & $13.8 \times 10^9$ & \cite{Godun14} \\
Oklo  &$ < 1.1 \times 10^{-8}$& $1.8 \times 10^9$&  \cite{DH15}\\
Quasars  &$ (-0.64\pm 0.65) \times10^{-6}$& $(8 - 11) \times 10^9$ & \cite{Martins17}\\
CMB  &$ (0.7\pm 2.5) \times10^{-3}$& $13.8 \times 10^9$ & \cite{HC18}\\
BBN  &$ (-2.2\pm 0.6) \times10^{-2}$& $13.8\times 10^{9}$ & \cite{MC17} \\
Supernovae type  Ia &$ (2.1\pm 0.8) \times 10^{-2}$&$ (0.1-9)\times 10^{10}$ & \cite{Kraiselburd2015} \\
\hline
White Dwarfs &$ (2.7\pm 9.1) \times10^{-5}$&  & \cite{Magano17} \\
\hline
\end{tabular}
\end{center}
\caption{Bounds on $\frac{\Delta \alpha}{\alpha}$. Columns: (1) Method considered (2) mean value and the corresponding $1-\sigma$ error, (3) the time interval for which the variation was measured, 4) Reference. The bound obtained from white dwarf data constraints the possible variation in $\alpha$ with the  gravitational potential.}
\label{tablealpha}
\end{table}

\begin{table}[!ht]
\begin{center}
\renewcommand{\arraystretch}{1.3}
\begin{tabular}{|l|l|l|l|}
\hline
Source & $\frac{\Delta \mu}{\mu}\pm \sigma$ & $\Delta t$ ({\rm yr}) & Reference  \\
\hline
Atomic Clocks  &$ (0.2 \pm 2.1) \times 10^{-16}$& $2 - 4$ & \cite{Godun14}\\
Quasars  &$ (-0.23\pm 0.1) \times10^{-6}$& $(6.4 - 12.4) \times 10^9$ & \cite{Martins17}\\
CMB  &$ (3.9\pm 7.4) \times10^{-3}$& $13.8 \times 10^9$ & \cite{HC18} \\
\hline
White Dwarfs & $ (-5.8\pm 4.1) \times 10^{-5}$& & \cite{Bagdonaite14}\\
\hline
\end{tabular}
\end{center}
\caption{Bounds on $\frac{\Delta \mu}{\mu}$. Columns: (1) Method considered (2) mean value and the corresponding $1-\sigma$ error, (3) the time interval for which the variation was measured, 4) Reference. The bound obtained from white dwarf data constraints the possible variation in $\mu$ with the  gravitational potential.}
\label{tablemu}
\end{table}

\begin{table}[!ht]
\begin{center}
\renewcommand{\arraystretch}{1.3}
\begin{tabular}{|l|l|l|l|}
\hline
Source & $\frac{\dot G_N}{G_N} \pm \sigma$ in units of  $10^{-12}$ & $\Delta t$ ({\rm yr}) & Reference \\
\hline
BBN &$ (0.05\pm 0.35)$ & $13.8 \times 10^{9}$ & \cite{copy04,Bambi05} \\
LLR  &$ (0.07 \pm 0.38) $&  $40$ &\cite{Hofmann10}\\
Supernovae  &$ < 10 $& $5 \times 10^9$ & \cite{Gazta02} \\
Helioseismology & $ < -1.6 $& $4.6 \times 10^{9}$ &\cite{Guenther98}\\
White Dwarfs & $ < -1.8 $&  $7.7 \times 10^9$ &\cite{Garciaberro11} \\
\hline 
\end{tabular}
\end{center}
\caption{Bounds on $\frac{\dot G_N}{G_N}$. Columns: (1) Method considered (2) mean value and the corresponding $1-\sigma$ error, (3) the time interval for which the variation was measured, 4) Reference. }
\label{tableG}
\end{table}

\section{Acknowledgments}

S.L. is grateful to IAU, IUPAP and the University of Leicester for financial support to attend the IAU 357 Symposium.
S.L. is also  supported  by the National Agency for the Promotion of Science and Technology (ANPCYT) of Argentina grant PICT-2016-0081; and by grant UBACYT 20020170100129BA (2018) from Buenos Aires University.

\end{document}